\documentclass{jetpl}
\usepackage{color,amsmath,amsfonts,graphics,epsfig,amssymb}
\usepackage[english]{babel}
\usepackage{hyperref}
\twocolumn
\lat

\newcommand{\black}{\color{black}}
\newcommand{\red}{\black}
\title{
\begin{flushright}
\normalsize \rm INR-TH-2023-024
\end{flushright}
\vspace{8mm}
Anomalous cosmic-ray correlations revisited with a complete full-sky sample of BL~Lac type objects
}

\rtitle{Cosmic rays and BL Lacs}

\sodtitle{Anomalous cosmic-ray correlations revisited with a complete full-sky sample of BL~Lac type objects}

\author{M.\,A.\,Kudenko\thanks{Corresponding author: {\tt kudenko.ma19@physics.msu.ru}} and S.\,V.\,Troitsky}

\rauthor{M.\,A.\,Kudenko and S.\,V.\,Troitsky}

\sodauthor{M.\,A.\,Kudenko, S.\,V.\,Troitsky}

\address{Institute for Nuclear Research of the Russian Academy of
Sciences,\\
60th October Anniversary prospect 7A, 117312 Moscow, Russia\\
and\\
Faculty of Physics, Lomonosov Moscow State University, 1-2 Leninskiye Gory, 119991 Moscow, Russia}

\dates{December 12, 2023}{January 15, 2024}

\abstract{Cosmic rays with energies above $10^{19}$~eV, observed in 1999--2004 by the High Resolution Fly's Eye (HiRes) experiment in the stereoscopic mode, were found to correlate with directions to distant BL~Lac type objects (BL Lacs), suggesting non-standard neutral particles travelling for cosmological distances without attenuation. This effect could not be tested by newer experiments because of their inferior angular resolution. The distribution in the sky of BL Lacs associated with cosmic rays was found to deviate from isotropy, which might give a clue to the interpretation of the observed anomaly. However, previous studies made use of a sample of BL Lacs which was anisotropic by itself, thus complicating these interpretations. Here, we use a recently compiled isotropic \red complete \black sample of BL Lacs and the same HiRes data to confirm the presence of correlations and to strengthen the case for the local large-scale structure pattern in the distribution of the correlated events in the sky. Further tests of the anomaly await new precise cosmic-ray data.
}
\begin{document}
\maketitle

\noindent\textbf{1. Introduction.} 
Two decades ago, puzzling correlations between arrival directions of ultra-high-energy cosmic rays, detected by the High Resolution Fly's Eye experiment (HiRes), and the BL Lac type objects (BL Lacs) were discovered \cite{HiRes:we}. These BL Lacs, which constitute a subclass of blazars, are active galactic nuclei with jets pointing to the observer, and are located at cosmological distances. Though they are plausible sites of cosmic-ray acceleration, charged nuclei would be deflected by cosmic magnetic fields, while the angular distances between arrival directions of HiRes events and these sources were significantly smaller than the expected deflections. However, neutral particles capable to reach the observer from such a large distance are missing in the Standard Model of particle physics \cite{TT:neutral}. This resulted in the obvious interest to the observation, which called to new-physics explanations.

Confirmed in Ref.~\cite{HiRes:HiRes}, where unpublished HiRes data were used, \red with the same pre-trial p-value of $2\times 10^{-4}$, corresponding to the post-trial $p\sim 10^{-3}$, \black the correlations \cite{HiRes:we} remain neither tested with other experiments nor unambiguously explained. The reason for the lack of tests is in the fact that none of modern experiments reaches the angular resolution of HiRes stereo, where, according to the Monte-Carlo simulations, 68\% of events were reconstructed within $0.55^\circ$ from their thrown direction, and 95\% -- within $1.26^\circ$ \cite{Finley:thesis}. Newer experiments increased the distance between detectors in order to cover larger effective area, hence worsened their angular resolution. \red The only \black attempt to test the results of \cite{HiRes:we,HiRes:HiRes} was reported at a conference \cite{Auger:BLL} but was not conclusive. \red Indeed, Ref.~\cite{BL-estimate} estimated the numbers of cosmic-ray events required to test the correlations \cite{HiRes:we} by different experiments. For the surface detector of the Pierre Auger Observatory, this number was about 3500, while only 1672 events were used in Ref.~\cite{Auger:BLL}.
\black See e.g.\ Refs.~\cite{FRT:2009,ST:pattern} for discussion \red of other complications related to these tests. \black 

This situation motivated one to use the original data to obtain maximun of information about these correlations. The clues to the interpretation might be found either from the astrophysical side (which BL Lacs are associated with cosmic rays, and how they are different from others), or from the cosmic-ray side (which cosmic-ray events are associated with BL Lacs, what could be said about their primary particle types or about the distribution of their arrival directions). 

The astrophysical approach was followed in Ref.~\cite{ST:index}, where it was found that the correlations are dominated by a particular class of BL Lacs selected by their broadband optical to X-ray spectral index, related to their physical properties. The information about the type of primary particles of correlated events cannot be obtained from public data, nor is discussed in Ref.~\cite{HiRes:HiRes}. The global distribution of arrival directions was studied in Ref.~\cite{ST:pattern}, and we return to it in the present work.

It is not easy to explain the observed correlations, even with the help of new physics. A promising possibility is provided \cite{FRT:2009} by axion-like particles (ALPs) with certain parameters, which mix with photons in the external magnetic field and travel unattenuated for cosmological distances, see e.g.\ Refs.~\cite{ST-mini-rev,Roncadelli-review2022} for reviews. Conversion of energetic photons to ALPs in the magnetic field of the source galaxy, with subsequent reconversion back to photons in the Milky-Way field, was proposed to explain observations of anomalous energetic gamma rays from blazars, see e.g.\  \cite{Serpico}, and a recent gamma-ray burst GRB~221009A, see e.g.\ \cite{ST-GRB-JETPL}. However, the conversions are suppressed in the galactic magnetic fields for ultra-high energies considered here, while weaker fields in clusters and filaments of the Large Scale Structure (LSS) might work \cite{FRT:2009}. A prediction of this scenario is thus the clustering of correlating events towards local LSS structure, which was indeed found in Ref.~\cite{ST:pattern}.

One complication of searches for the anisotropy of correlated events is the intrinsic anisotropy of the \red incomplete \black BL Lac catalog \cite{VCV2001} used in previous studies. Being a compilation of all data available for that moment, it covers different areas of the sky with different and not documented sensitivities. Besides the terrestrial reasons for anisotropy (different fields of view of telescopes), there is a significant Galactic anisotropy caused by the absorption near the Galactic plane (objects in Ref.~\cite{HiRes:we} were selected by their observed visual magnitude, not corrected for the absorption) and by consequent complications of spectral observations required to determine the BL Lac nature of the source. In general, these anisotropies have nothing to do with the LSS and can hardly affect the conclusions of the study, but biases could not be excluded. For instance, one of the suspects is the occasional almost perpendicular orientation of the Galactic and Supergalactic planes\red: because of this, the fraction of directions falling in the zone of avoidance is smaller for the local filament, compared to the full HiRes field of view\black.

With the purpose \red of avoiding \black such hard-to-control biases, we have recently compiled a new catalog of optically selected blazars \cite{iso-catalog} with the isotropic full-sky coverage. As a part of the project, a sample of confirmed BL Lacs was determined following the same criteria which were used in \cite{VCV2001}, see Ref.~\cite{Veron-criteria} for details. Here, we use this isotropic sample to test the LSS-related pattern \cite{ST:pattern} in the distribution of BL Lacs correlated with HiRes cosmic rays. We also make some comments regarding future tests of the correlations with new cosmic-ray data.

\vskip 1mm
\noindent\textbf{2. Cosmic-ray and BL Lac samples.}
For the present study, we use the same sample of arrival directions of 271 events detected by HiRes stereo \cite{HiRes:271events}, which was used in \cite{HiRes:we,ST:index,ST:pattern} and, with the addition of unpublished lower-energy events, in \cite{HiRes:HiRes}. Details on the sample may be found in Refs.~\cite{HiRes:271events,Finley:thesis}, but no information about individual events, except for the arrival directions published in the form of a postscript plot, is available.

The sample of BL Lacs we use is described in Ref.~\cite{iso-catalog}. It was constructed on the basis of full-sky catalogs of blazars selected by very-long-baseline radio interferometry (VLBI) and by gamma-ray observations. \red The sample is a combination of two complete samples, flux-limited at 8~GHz VLBI and above 1~GeV, respectively.\black

The selection of optically bright sources, \red motivated by the cuts used in Ref.~\cite{HiRes:we} and not related to the completeness of the isotropic sample\black, was performed on the basis of the GAIA DR3 $G$-band magnitude, corrected for the Galactic absorption, $G_{\rm corr}<18^{\rm m}$. There are 336 sources in the sample, and their distribution in the sky satisfies the quantitative criteria of isotropy described in Ref.~\cite{iso-catalog}.

For comparison, we also use the original sample of 156 confirmed BL Lacs selected from the catalog \cite{VCV2001} by the cut on their \textit{uncorrected} visual magnitude $V<18^{\rm m}$. The $V$ and $G$ bands correspond to similar wavelengths, though the latter one is wider.

\vskip 1mm
\noindent\textbf{3. HiRes cosmic rays correlated with the new BL-Lac sample.}

\vskip 0.5mm
\noindent{\sl 3.1. Directional correlations.}
We start with repeating the original analysis in exactly the same way as in Ref.~\cite{HiRes:we}, but with the new BL-Lac catalog. For a given angle $\theta$, one counts the number of pairs ``BL Lac -- HiRes arrival direction'' separated by the angle $\le\theta$. Then the same procedure is repeated for a large number of simulated sets of arrival directions, and the p-value measuring how often this or larger number of pairs can be observed by chance, is determined. Note that simulated arrival directions are not isotropic but follow the HiRes stereo exposure, see Ref.~\cite{HiRes:we} for details of the simulation. Like in \cite{HiRes:we}, we consider $0^\circ \le \theta \le 5^\circ$ in steps of $0.1^\circ$.

\begin{figure}
    \centering
    \includegraphics[width=\linewidth]{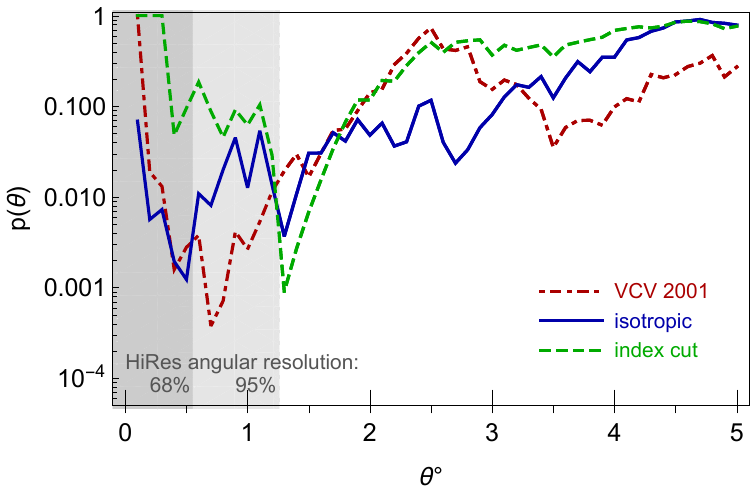}
    \caption{\label{fig:corr} \sl
\textbf{Figure~\ref{fig:corr}.}
Pre-trial p-value of random associations of HiRes stereo cosmic rays with three BL Lac samples, as a function of the searching cone opening $\theta$. The red dot-dashed line corresponds to the sample from \cite{VCV2001} used in \cite{HiRes:we}, the blue full line -- to the isotropic sample \cite{iso-catalog} used here, the green dashed line -- to the subsample selected by the optical to X-ray spectral index, see the text. Shading corresponds to the opening angles containing 68\% (dark) and 95\% (light) of HiRes events \cite{Finley:thesis}.
}
\end{figure}

The results of the pair-counting analysis are shown in Fig.~\ref{fig:corr}, 
where, for comparison, the results of \cite{HiRes:we} are also presented. We do not attempt to estimate the post-trial p-value here, because the two samples overlap heavily (52 out of 156 objects from the old sample are present in the new catalog), and the correlations have been already established in Refs.~\cite{HiRes:we,HiRes:HiRes}. Instead, we note that the pre-trial p-values for both samples are of the same order, despite the differences in catalogs, and that the minima correspond to slightly different values of $\theta$ in the range of 68\% to 95\% CL HiRes angular resolution. While for $\theta=0.8^\circ$, singled out in Ref.~\cite{HiRes:we}, 11 pairs are found in each of the two catalogs, these BL Lacs are not entirely the same: only 7 of 11 ones are present in both samples. At the same time, the background is larger in the new sample, because it contains more objects, and the p-values are correspondingly higher. This may be related to the correlations with LSS: the ``new'' BL Lacs in the isotropic catalog are located in the zones undercovered in the old one, notably around the Galactic plane. But, as we have pointed out, the Galactic plane, occasionally, is perpendicular to the local filament, so the fraction of correlated sources among these newly added sources is smaller compared to those in \cite{VCV2001}.

\vskip 0.5mm
\noindent{\sl 3.2. Optical to X-ray index selection.}
For completeness, we also perform a pair-counting test with the subsample of BL Lacs selected from the isotropic sample by the condition $\alpha_{\rm OX}<1$, where $\alpha_{OX}$ is the optical to X-ray spectral index defined and studied in Ref.~\cite{ST:index}. There are 73 such objects, and indeed, in accordance with \cite{ST:index}, the minimal p-value is lower for this sample, compared to the full one. This result is also presented in Fig.~\ref{fig:corr}. Note however that this subsample is not isotropic because of strongly nonuniform exposure of X-ray telescopes, as a result of which X-ray fluxes are not known for many of the sources in certain sky regions. 

\vskip 0.5mm
\noindent{\sl 3.3. Distribution of correlated directions in the sky.}
We turn now to the main topic of the present study and use the isotropic sample of BL Lacs to search for the LSS pattern in the distribution of correlated cosmic-ray arrival directions in the sky. Following Ref.~\cite{ST:pattern}, we use the weighted density of galaxies $f$ from the three-dimensional catalog of the Two Micron All Sky Survey \cite{2MRS-1,2MRS-2} as the LSS template. Details of the construction of the template, performed along the lines of Ref.~\cite{Koers:2008ba,Koers:2009pd}, are described in Ref.~\cite{ST:pattern}. \red Note that all BL Lacs are located far beyond the LSS, at distances between $\sim 150$~Mpc and a few gigaparsecs, while typical distances contributing to the template are $\lesssim 30$~Mpc. Correlations with sources located that far justifies the neglection of any attenuation at $\sim 30$~Mpc assumed in the calculation of $f$ in Ref.~\cite{ST:pattern}.\black 

\begin{figure}
    \centering
    \includegraphics[width=\linewidth]{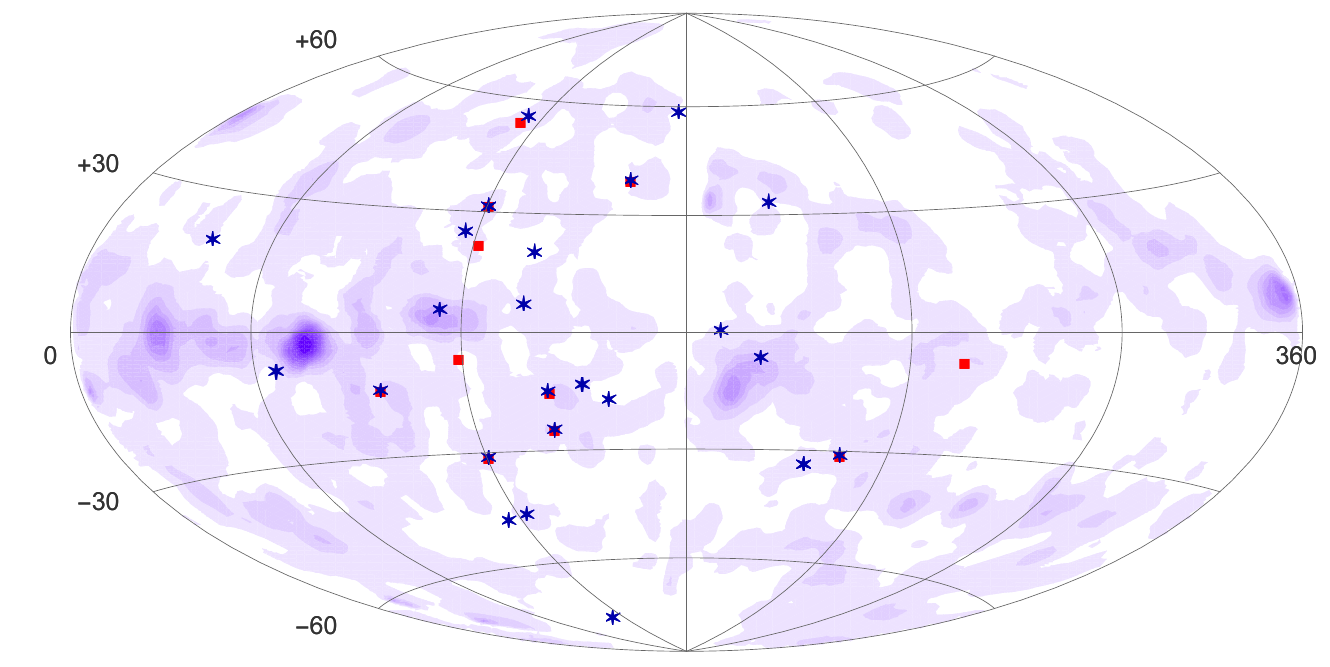}\\
    \caption{\label{fig:map2} \sl
\textbf{Figure~\ref{fig:map2}.}
Sky map with positions of BL Lacs associated with HiRes cosmic rays (supergalactic coordinates). Red boxes: the sample used in Ref.~\cite{HiRes:we}, $\theta=0.8^\circ$. Blue stars: the isotropic sample, $\theta=1.3^\circ$. Shading represents the weighted density of galaxies \cite{ST:pattern}.
}
\end{figure}

Figure~\ref{fig:map2}
presents the sky map with the correlating BL Lacs shown on top of this template distribution. One can see that, like in Ref.~\cite{ST:index}, correlating sources tend to fall in the darker shadow regions, corresponding to larger values of $f$. This is confirmed by a statistical study: making use of the Kolmogorov-Smirnov distances between sets of values of $f$ for correlated BL Lacs and for all HiRes events, we obtain the p-value of $3.9\times 10^{-3}$ for 19 BL Lac -- cosmic ray pairs\red, closer than the 95\% CL angular resolution, \black to follow the same distribution of $f$ as the entire HiRes data set. The same quantity for the sample used in \cite{HiRes:we} is $2.6\times 10^{-3}$. Therefore, despite the addition of many new sources to the catalog, the LSS pattern in the distribution of BL-Lac associated events remains the same. Note that this is true only for cosmic rays associated with BL Lacs: for the entire HiRes sample, correlations with LSS were not found \cite{HiRes-noLSS}.

\vskip 1mm
\noindent\textbf{4. Conclusions.}
We use the newly compiled isotropic full-sky sample of 336 confirmed optically selected BL Lacs to revisit the anomalous correlations of HiRes stereo cosmic rays with sources of this class. The correlations are found to be similar to the originally used, strongly anisotropic sample. The formal significance of the effect is not calculated for the new catalog because this study is not statistically independent from the original one. Pre-trial p-values are slightly higher than for the sample used in \cite{HiRes:we,HiRes:HiRes}, which fits well our expectations as discussed above. The condition for the optical to X-ray spectral index, $\alpha_{\rm OX}<1$, advocated in \cite{ST:index}, efficiently selects BL Lacs associated with cosmic rays in both samples. The main result, for which the isotropy of the sample is important, is the confirmation of the previously established \cite{ST:pattern} LSS pattern in the distribution of correlated events, which is not present for the entire cosmic-ray sample but is predicted by the ALP explanation of the anomaly.

While the low angular resolution of contemporary cosmic-ray observatories prevented one from testing the effects described here with new data, the large statistics collected by these experiments might help to partially overcome this difficulty. In the nearest future, tests of the BL Lac correlations with cosmic rays could be performed, in particular, with the help of the data collected by the Telescope Array experiment \cite{TelescopeArray:2008toq}, which is a successor of HiRes. \red To precisely evaluate the sensitivity of the expected analysis and refine the rough estimates of Ref.~\cite{BL-estimate}, dedicated simulations incorporating the experiment's specific details are essential.\black The present study allows us to make a few comments which might be useful in planning the strategy of these tests.
\begin{itemize}
\item
It is important to repeat the pair-counting test with the same set of BL Lacs which was used to establish the correlations \cite{HiRes:we} and with the same procedure. 
\item
The actual angle $\theta$, at which the correlations are most significant, depends on the fraction of cosmic-ray events coming from BL Lacs and on the fraction of BL Lacs in the sample, which emit cosmic rays. Both numbers are small and  fluctuate, so it makes sense to consider various values of $\theta$. 
\item 
To test the LSS pattern of correlated events, the use of the isotropic sample of sources is encouraged.
\item 
Given that the LSS pattern was predicted in the ALP scenario, and that another prediction in this case is that the primary particles are gamma rays, it would be important to explore possible photon content of the correlated events. However, other explanations might be possible, and any information about the primary particle types of the events would help to distinguish between the scenarios.
\end{itemize}
\vskip 1mm
\noindent\textbf{Acknowledgements.} The authors are indebted to M.~Kuznetsov and G.~Rubtsov for interesting and helpful discussions\red, and to three anonymous referees for useful comments\black. This work was supported by the Russian Science Foundation, grant 22-12-00253.

\bibliographystyle{nature1}
\bibliography{hires2023}
\end{document}